\documentclass[12pt]{article}
\usepackage{graphicx}
\input{epsf}

\newsavebox{\sboxpubnumber}
\newsavebox{\sboxpubdate}
\newcommand{\pubdate}[1]{\begin{lrbox}{\sboxpubdate}{#1}\end{lrbox}}
\newcommand{\pubnumber}[1]{\begin{lrbox}{\sboxpubnumber}{\begin{tabular}{l} #1
\\
				 \usebox{\sboxpubdate}
				 \end{tabular}}
                           \end{lrbox}
                           \pubblock}
\newcommand{\Title}[1]{\begin{center} {\Large #1 } \end{center}}
\newcommand{\Author}[1]{\begin{center}{ \sc #1} \end{center}}
\newcommand{\Address}[1]{\begin{center}{ \it #1} \end{center}}
\newcommand{\andauth}{\begin{center}{and} \end{center}}
\newcommand{\pubblock}{\rightline{
			\usebox{\sboxpubnumber}}}
\newenvironment{Abstract}{\begin{quotation}  }{\end{quotation}}
\newenvironment{Presented}{\begin{quotation} \begin{center}
             PRESENTED AT\end{center}\bigskip
      \begin{center}\begin{large}}{\end{large}\end{center}
      \end{quotation}}
\newcommand{\Acknowledgements}{\bigskip  \bigskip \begin{center} \begin{large}
             \bf ACKNOWLEDGEMENTS \end{large}\end{center}}

\def\to{\rightarrow}
\def\gev{\mbox{GeV}}
\def\ev{\mbox{eV}}
\def\mev{\mbox{MeV}}

\def\mpc{\mbox{Mpc}}

\def\vev#1{\langle {#1}\rangle}

\def\frac#1#2{{\textstyle{{#1}\over {#2}}}}

\def\lsim{\mathrel{\rlap{\lower4pt\hbox{\hskip1pt$\sim$}}
    \raise1pt\hbox{$<$}}}
\def\gsim{\mathrel{\rlap{\lower4pt\hbox{\hskip1pt$\sim$}}
    \raise1pt\hbox{$>$}}}
\def\sqr#1#2{{\vcenter{\vbox{\hrule height.#2pt
         \hbox{\vrule width.#2pt height#1pt \kern#1pt
         \vrule width.#2pt}
         \hrule height.#2pt}}}}

\newcommand{\beq}{\begin{equation}}
\newcommand{\eeq}{\end{equation}}
\newcommand{\bea}{\begin{eqnarray}}
\newcommand{\eea}{\end{eqnarray}}

\begin{document}

\begin{titlepage}
\pubdate{\today}                    
\pubnumber{DF/IST-12.2001} 

\vfill
\Title{Cosmological Bounds on an Invisibly Decaying Higgs Boson}
\vfill
\Author{O. Bertolami\footnote{E-mail: orfeu@cosmos.ist.utl.pt}}
\Address{Departamento de F\'\i sica, Instituto Superior T\'ecnico\\ 
Av. Rovisco Pais 1, 1049-001 Lisboa. Portugal}

\vfill
\andauth
\vfill
\Author{M.C. Bento}
\Address{Departamento de F\'\i sica, Instituto Superior T\'ecnico\\ 
Av. Rovisco Pais 1, 1049-001 Lisboa. Portugal}

\vfill
\andauth
\vfill
\Author{R. Rosenfeld}
\Address{Instituto de F\'\i sica Te\'orica\\
R.\ Pamplona 145, 01405-900 S\~ao Paulo - SP, Brazil}

\vfill
\begin{Abstract}
We derive bounds on the Higgs boson coupling $g^{\prime}$ to a stable 
light scalar which is regarded as a collisional dark matter candidate. 
We study the behaviour of this scalar, that we refer to as phion ($\phi$), 
in the early Universe for
different ranges of its mass. We find that a phion in the mass range of 
$100 ~\mev$ is excluded, while if its mass is about $1 ~\gev$,
a rather large coupling constant, $g^{\prime} \gsim 2$, and
$m_h \lsim 130~\gev$ are required in order to avoid overabundance. 
In the latter case, the invisible decay mode of the Higgs boson is 
dominant. 
\end{Abstract}
\vfill
\begin{Presented}
    COSMO-01 \\
    Rovaniemi, Finland, \\
    August 29 -- September 4, 2001
\end{Presented}
\vfill
\end{titlepage}
\def\thefootnote{\fnsymbol{footnote}}
\setcounter{footnote}{0}

\section{Introduction}

We have recently suggested that a light, stable, strongly self-coupled 
scalar field coupled with the Higgs field would be an interesting candidate 
for solving the problems of the Cold Dark Matter (CDM) model  
at galactic scales \cite{Bento1}. In here we discuss some early Universe 
history of this particle \cite{Bento2}. 
Our proposal involves a particle physics-motivated model, 
where the DM particles are allowed to self-interact so as to have a large 
scattering 
cross section and negligible annihilation or dissipation. The self-interaction
results in a characteristic length scale given by the mean free 
path of the particle in the halo.
This idea was originally proposed to suppress small scale power 
in the standard CDM model  
\cite{Carlson,Laix} and has been recently revived, in a general context, 
in order to address CDM difficulties at galactic scale \cite{Spergel}. 
Our model \cite{Bento1} is 
a concrete realization of this idea and involves an extra gauge singlet 
as the self-interacting, non-dissipative cold 
dark matter particle. Following Ref. \cite{Binoth}, we call this scalar 
particle phion, $\phi$, and assume that it couples
to the Standard Model (SM) Higgs boson, $h$, with a Lagrangian 
density given by:

\begin{equation}
{\cal L} = {1 \over 2} (\partial_\mu \phi)^2 - {1 \over 2} m_\phi^2 \phi^2  
- {g \over 4!} \phi^4 + g^{\prime} v \phi^2 h 
\quad,
\label{1}
\end{equation}%
where $g$ is the phion
self-coupling constant, $m_\phi$ its mass, $v=246\ \gev$ is the Higgs 
vacuum expectation value and $g'$ is the coupling between
$\phi$ and $h$. A model along these lines 
have been previously discussed \cite{Silveira}. Clearly the interaction term 
between the phion and the Higgs boson arises from a quartic interaction 
${g^{\prime} \over 2}  \phi^2 H^2$, where $H$ is the electroweak Higgs doublet.
As shown in \cite{Bento1}, the $\phi$ mass does not arise 
from spontaneous symmetry breaking since this would yield a tiny scalar 
self-coupling constant. The phion mass in (\ref{1}) should be regarded as 
as a phenomenological parameter arising from a more encompassing theory.
    
As is well known, scalar particles 
have been repeatedly invoked as DM
candidates \cite{Gradwohl,McDonald,Bertolami,Peebles,Goodman,Matos}; 
however, our 
proposal has the salient feature that it brings about a 
connection with the SM Higgs boson which could arise in extensions of the
SM. For instance, the hidden sector of heterotic string theories does 
give rise to 
astrophysically interesting self-interacting scalars \cite{Faraggi}. 
For reasonable values of $g^{\prime}$, the new scalar
would introduce a novel invisible
decay mode for the Higgs boson. This could, in principle, provide an
explanation for the failure in finding the Higgs boson at accelerators
sofar\cite{Bij}.

On the astrophysical front, recent observational data on large scale 
structure, Cosmic Microwave Background 
anisotropies and type Ia Supernovae suggest that $\Omega_{tot} \approx 1$, 
of which $\Omega_{baryons} \approx 0.05$ and $\Omega_\Lambda \approx 0.65$ 
\cite{Bahcall}; the remaining contribution, $\Omega_{DM}\approx 0.3$ 
(apart from neutrinos that may contribute a small fraction), comes from 
dark matter (DM), which
determines the hierarchy of the structure formation in the Universe. 
The most prominent theories of structure formation are now $\Lambda$CDM 
and QCDM, which consist, respectively, of the standard 
Cold Dark Matter (CDM) model 
supplemented by a cosmological constant or a dark energy, i.e. a negative 
pressure component.

In the CDM model, initial
Gaussian density fluctuations, mostly in non-relativistic collisionless
particles, the so-called cold dark matter, grow during the  inflationary 
period of the Universe and evolve, via  gravitational instability, into the 
structures one observes at present.
However, it has been found that the CDM model cannot sucessfully accomodate 
the data observed  on all scales. For instance, N-body simulations 
predict a number of halos 
which is about an order of magnitude grater than the 
observed number at the level of Local Group \cite{Mooreetal2,Klypin}.
Furthermore, astrophysical systems that are DM dominated, e.g.  
dwarf galaxies \cite{Moore,Flores-Primack,Burkert1} 
show shallow matter--density profiles with 
finite central densities. This contradicts high resolution N-body simulations 
\cite{Navarro,Ghigna,Mooreetal}, which have singular cores, 
with $\rho \sim r^{-\gamma}$ and $\gamma$ in the range between 1 and 2.
This can be interpreted as an indication of the fact that since 
cold collisionless DM particles do not have 
any characteristic length scale they lead, due to hierarchical 
gravitational collapse, to very dense dark matter halos that present 
negligible core radius.

On the other hand, recent numerical simulations   
\cite{Hannestad,Moore1,Yoshida,Wandelt} indicate that the self-interaction
of DM particles does bring noticeable 
improvements on the properties of the CDM model on small scales. 

At present, $\phi$ particles are non-relativistic, with typical 
velocities 
$v \simeq 200 $ km s$^{-1}$, and, therefore, it is impossible to dissipate
energy creating more particles in reactions like 
$\phi \phi \rightarrow \phi \phi \phi
\phi$.  Thus, as only the elastic channel is 
kinematically accessible, near 
threshold, the cross section is given by:

\begin{equation}
\sigma (\phi \phi \rightarrow \phi \phi) \equiv \sigma_{\phi \phi} 
= {g^2 \over 16 \pi s} \simeq {g^2 \over 64 \pi m_\phi^2}
\quad.
\label{eq:cross1} 
\end{equation}

A limit on $m_\phi$ and $g$ can be obtained by demanding that the mean free
path of the particle $\phi$, $\lambda_\phi$,  is in the interval 
$1~\mbox{kpc} < \lambda_{\phi} < 1~$Mpc \cite{Spergel}.  
Hence, we have:

\begin{equation}
\lambda_{\phi} = {1 \over \sigma_{\phi \phi} n_\phi} 
= {m_\phi \over \sigma_{\phi\phi} \rho^{h}_\phi} 
\quad,
\label{eq:cross11} 
\end{equation}
where $n_\phi$ and $\rho^{h}_\phi$ are, respectively, 
the number and mass density of  $\phi$ particles in the halo. 
Eqs. (\ref{eq:cross1}) and (\ref{eq:cross11}) imply that

\begin{eqnarray}
\sigma_{\phi \phi} & = & 2.1 \times 10^{3} 
\left({m_\phi \over \gev}\right) 
\left({\lambda_\phi \over \mpc} \right)^{-1}\nonumber\\ 
&\times & \left({\rho^{h}_\phi \over 0.4~\mbox{\gev cm}^{-3}} \right)^{-1}
~~\gev^{-2}
\quad,
\label{eq:cross2}
\end{eqnarray} 
which, in turn, leads to:

\begin{equation}
m_\phi= 13~g^{2/3}
\left({\lambda_{\phi} \over \mpc} \right)^{1/3} 
\left({\rho^{h}_\phi \over 0.4~\mbox{\gev cm}^{-3}}\right)^{1/3}\mev 
\quad.
\label{mass}
\end{equation}

We shall next analyse how the requirement that $\Omega_{\phi} h^2 
\simeq 0.3$, i.e. that the phion is a suitable DM candidate, and
that the phion is able to explain small scale
structure, leads to bounds on the couplings $g$ and $g^{\prime}$. 

\section{Phion density estimate}
\label{sec:estimate}

If the coupling constant $g^{\prime}$ is sufficiently small, 
phions decouple early in the 
thermal history of the Universe and are diluted by subsequent entropy 
production.
In Ref. \cite{Bento1}, it was considered out-of-equilibrium phion production 
via inflaton 
decay in the context of $N=1$ Supergravity inflationary models (see. 
e.g. \cite{Bento3} and references therein) and found that 
$\Omega_{\phi} h^2 \simeq 0.3$ can be naturally achieved.

On the other hand, for certain values of the coupling 
$g^\prime$, it is possible
that $\phi$ particles are in thermal equilibrium with ordinary matter. 
In order to determine whether this is the case, we will make the usual 
comparison between the thermalization rate $\Gamma_{th}$ and the expansion 
rate of the Universe $H$.

The thermalization rate is given by
\begin{equation}
\Gamma_{th} = n <\sigma_{ann} v_{rel}>
\quad,
\end{equation}
where 
$n = 1.2 \times T^3/\pi^2$  is 
the density of relativistic phions and $< \sigma_{ann} v_{rel}> $ is the 
annihilation
cross section averaged over relative velocities. 
On the other hand,  the expansion rate is given by:
\begin{equation}
H = \left({4 \pi^3 g_\ast \over 45}\right)^{1/2} {T^2 \over M_{P}} = 
1.66 \times g_\ast^{1/2} {T^2 \over M_{P}}
\quad.
\label{hubble}
\end{equation}

At temperatures above the electroweak phase transition, 
$T_{EW} \simeq 300~\gev$, a typical value in many extensions of the SM where 
one hopes to find the required features to achieve successful baryogenesis, 
the
order parameter (the vacuum expectation value of the Higgs field) vanishes, 
and hence the $\phi\phi h$ coupling is non-operative. 
However, this interaction term has its origin in the 
4-point coupling, $\phi \phi h h$, which can keep, at high temperatures, the
phion-Higgs system into thermal equilibrium.
Using the temperature as the center-of-mass energy, the cross section is given
by:
\begin{equation}
\sigma_{ann} v_{rel} \simeq {g^{\prime 2} \over 32 \pi T^2}
\quad,
\end{equation} 
which implies that phions are in thermal equilibrium for temperatures smaller
than 
\begin{equation}
T_{eq} \simeq {g^{\prime 2} M_P \over 32 \pi^3 g_\ast^{1/2}}
\quad. 
\end{equation}
Therefore,  phions cannot be in thermal equilibrium
before the
electroweak phase transition if $g^{\prime} \lsim 10^{-7}$. 

Thermal equilibrium can be achieved just below $T_{EW}$, when 
the trilinear coupling is operative, if $g^\prime$ is such that
the thermalization rate, $\Gamma_{th}$, exceeds the 
Hubble expansion rate. Let us quantify the conditions on $g^\prime$ to
satisfy this condition.

The phion annihilation cross section ($T \gsim m_h$) 
is given by the relativistic Breit-Wigner resonance formula:

\begin{equation}
\sigma_{ann} v_{rel} = {
4 \pi (s/m_h^2) \Gamma(h \rightarrow \phi \phi) \Gamma_h
\over
(s - m_h^2)^2 +
m_h^2\Gamma_h^2}
\quad,
\label{sigma}
\end{equation}
where $\Gamma_h$ is the total Higgs decay rate. At the resonance peak 
($s = m_h^2$) it simplifies to
\begin{equation}
\sigma_{ann} v_{rel} = {4 \pi \over m_h^2} BR(h \rightarrow \phi \phi)
\quad.
\end{equation}

From the Higgs decay width into phions
\begin{equation}
\Gamma(h\rightarrow \phi\phi)={g{^\prime}^{2} v^{2} 
(m_h^2 - 4  m_{\phi}^2)^{1/2} \over 32 \pi m_h^2}
\quad,
\label{eq:gphi}
\end{equation}
we get the decoupling temperature in the limit $m_h \gg m_\phi$:

\begin{equation}
T_D \simeq 150 {\Gamma_h m_h^3 \over g^{\prime 2} M_P v^2} 
\quad.
\end{equation}
Thus, in order to have 
a decoupling temperature of the order of the
Higgs mass, the coupling constant should be fairly small:
\begin{equation}
g^{\prime} \simeq 10^{-10}
\quad,
\end{equation}
where we have introduced the SM value 
of $\Gamma_h = 3.2 ~\mev$, obtained from
the code HDECAY \cite{hdecay}, for
a $m_h=115 ~\gev$ Higgs boson.

Hence, if $g^{\prime} \geq 10^{-10}$, the phions will be kept 
into thermal equilibrium after the electroweak phase transition.
In this situation, there are two possible scenarios depending whether
they decouple while relativistic or otherwise.

\begin{figure}
\centerline{\epsfysize=5cm \epsfbox{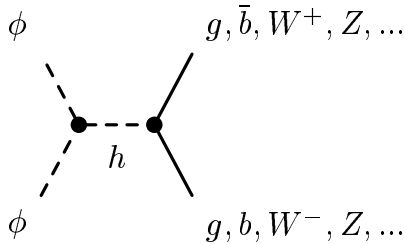}}
\caption{Feynman diagramm for phion annihilation via Higgs exchange.}
\label{fig:phiong}
\end{figure}

In order to study these scenarios, we have to establish the
decoupling temperature at $T \simeq m_\phi \ll T_{EW}$, in which case the 
phion annihilation cross section involves virtual Higgs exchange 
($h^{\ast}$), as in Figure~\ref{fig:phiong}, and is given 
by \cite{Burgess}:

\begin{equation}
\sigma_{ann} v_{rel} ={8 {g^{\prime}}^2 v^2 \over (4 m_\phi^2-m_h^2)^2 +
m_h^2\Gamma_h^2} F_X
\quad,
\label{sigma}
\end{equation}
where

\begin{equation}
F_X=\lim_{m_{h^\ast} \rightarrow 2 m_\phi}
\left({\Gamma_{h^{\ast} X} \over m_{h^\ast}}\right) 
\quad,
\label{fx}
\end{equation}
and $\Gamma_{h^\ast X}$
refers to the width for the decay $h^\ast \rightarrow X$ ($ X \neq \phi \phi$,
since we are dealing with inelastic scattering only), for
$m_{h^\ast} = 2 m_{\phi}$. For the mass range of interest to us, 
$m_\phi \sim 10-100~\mev$, one finds $F_X \sim 10^{-13}$ \cite{Gunion}.

Under these conditions, the relationship between the coupling 
constant $g^{\prime}$ and the decoupling temperature $T_D$ is given by 

\begin{equation}
g^{\prime 2} = 5.5 {(m_h/100 \gev)^4 \over (T_D/\mev)}
\quad.
\end{equation}

If $g^{\prime} \leq 0.1$, the phions decouple while relativistic and are 
as abundant as photons. 
Since we are interested in stable light phions, 
it is a major concern avoiding phion 
overproduction if it decouples while relativistic.
Actually, we find that there is  an analogue of Lee-Weinberg limit for
neutrinos (see e.g. \cite{Kolb}):

\begin{equation}
\Omega_\phi h^2 \simeq 0.08 {m_\phi \over 1 ~ \ev}
\quad,
\end{equation}
yielding a very stringent bound, $m_\phi \lsim 4 ~\ev$, and therefore,  
$g \lsim 2.5 \times 10^{-10}$, for the phion self-coupling constant 
so to solve the small scale structure problem of the
collisionless CDM.

In order to ensure that the phions decouple non-relativistically and 
that their
abundance reduces to acceptable levels without fine-tuning the
self-coupling constant, $g^{\prime} > 0.1$ is required. It follows from  
standard methods that
the phion relic abundance \cite{Burgess,Kolb}, is given by

\begin{equation}
\Omega_{\phi} h^2 = {1.07 \times 10^9 x_F \over g_{\ast}^{1/2} M_P 
\vev{\sigma_{ann} v_{rel}}}
\quad,
\label{omega}
\end{equation}
where $g_{\ast}$ denotes the number of degrees of freedom in equilibrium at
annihilation and $x_F \equiv m_{\phi}/T_F$ is the inverse of the freeze-out 
temperature in units of the phion mass. 
The relevant cross section is the phion annihilation 
cross section involving virtual Higgs exchange, Eq. (\ref{sigma}), with 
$F_X \sim 10^{-13}$ \cite{Gunion}. The freeze-out temperature is set by 
the solution of the Boltzmann equation

\begin{equation}
x_F \simeq \ln[0.038 (g_{\ast} x_F)^{-1/2} M_P m_{\phi}
\vev{\sigma_{ann} v_{rel}}]
\quad.
\label{xf}
\end{equation}

For obtaining $x_F \geq 1$ and to apply Eq. (\ref{omega}), it is required that 
$g^{\prime} \gsim 1.2$, for $m_\phi = 50 ~\mev$ and $m_h = 115 ~ \gev$. 
However, due
to the smallness of the cross section, the relic abundance of the phion is
several orders of magnitude larger than the observed value. Therefore, one can
conclude that a phion particle with a mass in this range is excluded.

\begin{figure}
\centerline{\epsfysize=7cm \epsfbox{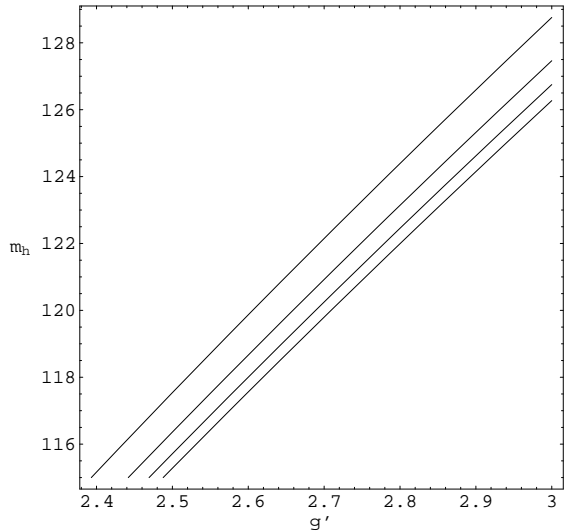}}
\caption{Contour of $\Omega_\phi h^2=0.3$ as a function of $m_h$ (in GeV) and  
$g^\prime$, for $m_{\phi}=0.5~\gev$ (top),~$1.0,~1.5$ and $2~\gev$ (bottom).}
\label{fig:mphiall}
\end{figure}

The smallness of the phion annihilation cross section for $m_\phi \simeq 50
~\mev$ has its origin in the small factor 
$F_X \simeq 10^{-13}$. However, this factor
increases significantly with larger phion masses. For 
$m_\phi \simeq 1 ~\gev$, $F_X \simeq 10^{-7}$.
We find that the requirement $\Omega_{\phi} h^2 \simeq 0.3$ implies that 
$m_{\phi} \gsim 500~\mev$ and $g^{\prime} \gsim 2$ (which is at the edge of 
validity of perturbation theory), a solution which holds 
only for $m_h \lsim 130~\gev$. Heavier phion and Higgs particles tend to make 
$\Omega_{\phi} h^2 > 0.3$. 
Our results are depicted in Figure \ref{fig:mphiall}.

For these large values of the coupling constant, the decay width of the Higgs
into phions is given by:
\begin{equation}
\Gamma(h\rightarrow \phi\phi)= 5.23~\left({m_h \over 115~\gev}\right)^{-1} 
g{^\prime}^{2}
~ \gev.
\end{equation}

Thus, the Higgs width is totally dominated by the 
invisible decay mode and this model can be easily tested at future 
colliders.


\section{Conclusions}

In this contribution, we have derived, bounds on $g^{\prime}$, the 
coupling constant of the Higgs boson to a stable scalar particle,
which contribute to Higgs decay 
via invisible channels. This particle, the phion, is suitable 
self-interacting dark matter candidate 
and allows for a solution of 
the difficulties of the CDM model on small scales. 

We find that, for $g^{\prime} \lsim 10^{-10}$, the phions never reach 
thermal equilibrium and are only produced by 
out-of-equilibrium decay of the inflaton field \cite{Bento1}. In this 
scenario, the
phion does not contribute to the invisible Higgs boson decay channel.
For $g^{\prime} \gsim 10^{-10}$, we have found that, if 
$g^{\prime} \lsim 0.1$, the phion decouples while still 
relativistic and a limit for its mass, $m_\phi \lsim
4~\ev$ can be derived, which, in turn, implies in a strong bound on 
the phion self-coupling constant, $g \lsim 10^{-9}$. 
On the other hand, if $g^{\prime} \gsim 1$,
the phion decouples while non-relativistic; but, its abundance is 
not cosmologically acceptable for phion masses in the range of $50 - 100
~\mev$ due to the small annihilation cross section. For masses in the range of
$0.5 - 2 ~\gev$, we find that abundances of $\Omega_\phi h^2 \simeq 0.3$
require large values of the coupling $g^{\prime} \simeq 2.5$ and 
$m_h \lsim 130~\gev$. In this scenario,
the Higgs width is dominated by the invisible $h \to \phi \phi$ mode and can be
tested at future colliders.

\Acknowledgements

\noindent
M.C.B. and O.B. would like to acknowledge the partial support 
of FCT (Portugal) under the grant POCTI/1999/FIS/36285. R.R. is partially 
supported by a PRONEX (CNPq - Brazil) grant.


\end{document}